\documentclass[letterpaper]{article} % DO NOT CHANGE THIS
\usepackage[preprint]{aaai2027}  % Public preprint mode; do not modify aaai2027.sty
\usepackage{amsmath}
\usepackage{amssymb}   % for \checkmark
\usepackage{booktabs}  % for professional tables
\usepackage{graphicx}  % for figures
\usepackage{multirow}  % for multirow table cells
\usepackage{subcaption}% for RQ1 subfigures
\usepackage[hyphens]{url}  % DO NOT CHANGE THIS
\usepackage{natbib}  % DO NOT CHANGE THIS AND DO NOT ADD ANY OPTIONS TO IT
\usepackage{adjustbox}
\usepackage{caption} % DO NOT CHANGE THIS AND DO NOT ADD ANY OPTIONS TO IT

\usepackage{fancyhdr}

\fancypagestyle{preprintfirstpage}{
  \fancyhf{}

  \fancyfoot[C]{\small\itshape Preprint version.}
}

\title{Beyond ``What to Retrieve'': Uncertainty in Retrieval-Augmented Code Generation}

\author{
Chandan Kumar Sah,
Li Zhang,
Xiaoli Lian
}
\affiliations{
Beihang University, Beijing, China\\
\texttt{sahchandan98@buaa.edu.cn}\\
}

\begin{document}
\maketitle
\thispagestyle{preprintfirstpage}

\begin{abstract}
Repository-level code generation relies on heterogeneous evidence whose
relevance, compatibility, and completeness are inherently uncertain.
Similar-code examples, repository context, and project-specific APIs may
provide complementary information, but can also introduce noisy, redundant,
or conflicting signals. Existing retrieval-augmented approaches primarily
optimize retrieval relevance without explicitly modeling how uncertainty in
retrieved evidence affects downstream generation. We introduce \textsc{OpenCoder}, an uncertainty-aware framework that estimates
source-specific uncertainty, uses it to filter and rank heterogeneous evidence,
and guides generation, verification, and repair. A factorial analysis over API knowledge, repository
context, and similar-code evidence reveals no universal additive source
ranking; instead, significant cross-source interactions depend on the
accompanying evidence and LLM backend. On an expanded 32-task
RepoExec-inline evaluation, \textsc{OpenCoder} improves GPT selected-output
correctness over Baseline RAG from 56.25\% to 78.13\%. However, it matches a
verification-and-repair control, and the corresponding Gemini improvement is
not statistically supported, indicating backend-dependent benefits.
Target-aware API refinement also substantially improves API-set retrieval.
These findings support treating uncertainty as an actionable control signal
for repository-level retrieval, verification, and repair.
\end{abstract}

\noindent
\small
Our code and supporting materials are available at\\
\url{https://github.com/Rocky5502/OpenCoder_V1}.
\normalsize

\section{Introduction}
\label{sec:introduction}
Large language models (LLMs) have substantially advanced automated code
generation, demonstrating strong performance on function-level programming
tasks~\cite{chen2021evaluating,roziere2023codellama,guo2024deepseekcoder}.
However, real-world software development rarely involves isolated functions.
Repository-level code generation requires models to understand cross-file
dependencies, project conventions, private APIs, and execution environments
that may not be captured by parametric knowledge or the local code
context~\cite{yu2023codereval,hai2024repoexec,yang2024execrepobench}. These
requirements make repository-level generation particularly challenging:
relevant information is distributed across large codebases, while the context
available to an LLM remains limited.
Retrieval-augmented generation (RAG) addresses this challenge by supplying
models with external repository knowledge~\cite{lewis2020rag}. Existing
repository-level methods retrieve similar code, cross-file context, dependency
information, or project-specific APIs. RepoCoder alternates retrieval and
generation, RepoFormer selectively invokes retrieval, GraphCoder exploits
structured code-context graphs, and RLCoder learns retrieval policies from
generation feedback~\cite{zhang2023repocoder,wu2024repoformer,
liu2024graphcoder,wang2024rlcoder}. More recent work combines heterogeneous
evidence sources and shows that repository context and API knowledge can be
more useful than naively retrieved similar code~\cite{gu2025what}. Collectively,
these studies demonstrate the value of repository-aware retrieval, but they
primarily optimize \emph{what} information should be retrieved.

\begin{figure*}[t]
    \centering
    \includegraphics[width=0.96\textwidth]{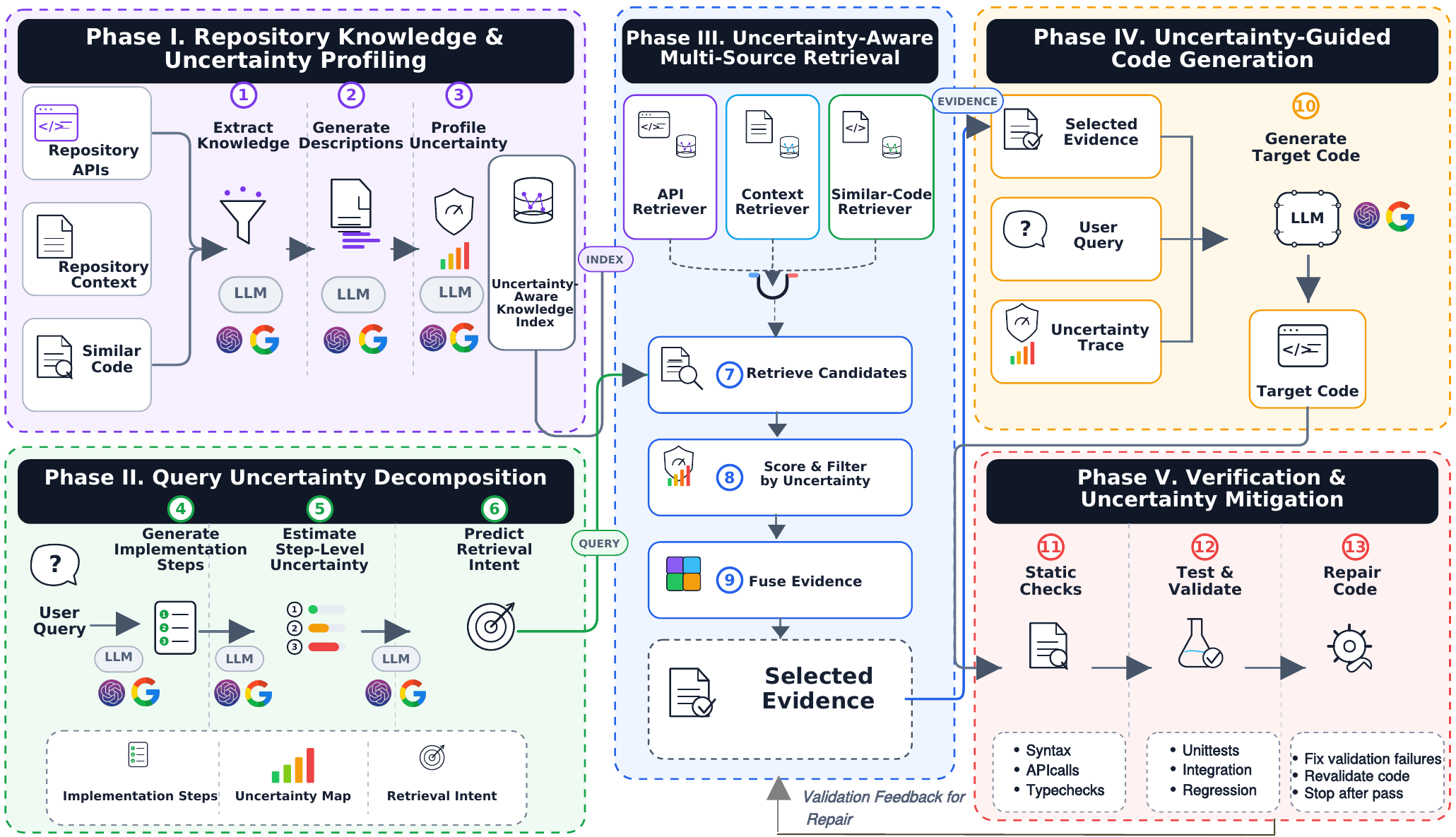}
    \vspace{-5pt}
    \caption{Overview of \textsc{OpenCoder}'s uncertainty-aware pipeline for
repository indexing, evidence retrieval and fusion, code generation,
verification, and repair.}
\label{fig:opencoder-framework}
\end{figure*}

Retrieval relevance alone does not ensure reliable generation. Similar-code
evidence may be semantically related yet functionally incompatible with the
target task, while repository context may contain redundant or conflicting
dependencies. API evidence may be incomplete, incorrectly scoped, or
incompatible with the target implementation. This is particularly
consequential because executable LLM-generated code can still contain
substantial API misuse~\cite{zhong2024robustapi}. Moreover, combining additional
evidence can increase prompt noise and propagate retrieval errors into
generation. Related RAG research has explored selective retrieval,
self-reflection, corrective retrieval, and source-reliability estimation
~\cite{asai2023selfrag,yan2024crag,hwang2024rarag}, but these mechanisms have
not been systematically developed for heterogeneous repository evidence and
execution-based code generation. As summarized in
Table~\ref{tab:retrieval_evidence}, existing approaches use different
combinations of similar code, repository context, and API knowledge, whereas
OpenCoder additionally models the uncertainty associated with these evidence
sources. To address this gap, we introduce \textsc{OpenCoder}, an uncertainty-aware
framework for retrieval-augmented repository-level code generation. As
illustrated in Figure~\ref{fig:opencoder-framework}, OpenCoder constructs a
repository knowledge index, decomposes a query into implementation steps, and
retrieves evidence from similar code, repository context, and project-specific
APIs. It estimates source-specific uncertainty and integrates evidence according to
retrieval relevance, predicted source utility, and evidence uncertainty. The resulting uncertainty trace guides evidence filtering and generation,
while executable validation controls final candidate selection and repair. OpenCoder therefore treats uncertainty as an actionable
decision signal rather than only a post-hoc confidence score.
\begin{table}[t]
\centering
\scriptsize
\setlength{\tabcolsep}{2.7pt}  % Slightly increased for better readability

\begin{tabular}{@{}lcccc@{}}
\toprule
\textbf{Approach} 
& \textbf{Similar-Code} 
& \textbf{API} 
& \textbf{Repo Context}
& \textbf{Uncertainty} \\
\midrule
A$^3$-CodGen~\citep{liao2024a3codgen}
& $\times$ & $\checkmark$ & $\checkmark$ & $\times$ \\
RepoCoder~\citep{zhang2023repocoder}
& $\checkmark$ & $\times$ & $\checkmark$ & $\times$ \\
RepoFormer~\citep{wu2024repoformer}
& $\checkmark$ & $\times$ & $\checkmark$ & $\times$ \\
RepoMinCoder~\citep{li2024repomincoder}
& $\checkmark$ & $\times$ & $\checkmark$ & $\times$ \\
RLCoder~\citep{wang2024rlcoder}
& $\checkmark$ & $\times$ & $\checkmark$ & $\times$ \\
R$^2$C$^2$-Coder~\citep{deng2024r2c2coder}
& $\checkmark$ & $\times$ & $\checkmark$ & $\times$ \\
GraphCoder~\citep{liu2024graphcoder}
& $\checkmark$ & $\times$ & $\checkmark$ & $\times$ \\
RepoFuse~\citep{liang2024repofuse}
& $\checkmark$ & $\checkmark$ & $\checkmark$ & $\times$ \\
AllianceCoder~\citep{gu2025what}
& $\checkmark$ & $\checkmark$ & $\checkmark$ & $\times$ \\
\midrule
\textbf{OpenCoder} (Ours)
& $\checkmark$ & $\checkmark$ & $\checkmark$ & \textbf{$\checkmark$} \\
\bottomrule
\end{tabular}
\vspace{-5pt}
\caption{Retrieval evidence and uncertainty treatment in representative
repository-level code-generation methods. OpenCoder jointly models similar
code, repository context, API knowledge, and source-specific uncertainty.}
\label{tab:retrieval_evidence}
\end{table}

We evaluate \textsc{OpenCoder} using GPT and Gemini backends under a frozen,
matched five-candidate protocol. Before inspecting method outputs, we audit 25
additional RepoExec tasks, retain 18 that satisfy the executable protocol, and
expand RepoExec-inline from 14 to 32 matched tasks. With GPT,
\textsc{OpenCoder} improves selected-output correctness over Baseline RAG from
56.25\% to 78.13\% (a 21.88-point gain; 95\% CI [6.25,40.63]; nominal
McNemar $p=.039$), while tying the matched \emph{RAG + Verify/Repair} control.
With Gemini, selected-output and candidate-set differences are not statistically
supported, and the control is strongest at most metrics. All paired
RepoExec-inline Pass@\(k\) intervals include zero. On context-limited
ExecRepoBench, the control substantially outperforms \textsc{OpenCoder},
revealing a boundary condition under incomplete repository evidence.
Target-aware API refinement nevertheless increases macro API F1 from 43.4\%
to 64.8\% for GPT and to 59.1\% for Gemini. \noindent\textbf{Our main contributions are:}
\begin{itemize}
    \item We formulate retrieval-augmented repository-level code generation as
    decision-making under uncertainty over heterogeneous repository evidence.

    \item We introduce \textsc{OpenCoder}, a unified framework that estimates
    source-specific uncertainty and operationalizes it across evidence
    filtering, uncertainty-aware multi-source integration, generation, verification, and repair.

    \item Through a full factorial analysis of API knowledge, repository
context, and similar-code evidence, we show that retrieval utility is
interaction-dependent rather than governed by a fixed source ranking,
empirically motivating uncertainty-aware evidence integration.

    \item We conduct a controlled matched evaluation across GPT and Gemini on
    executable repository-level tasks. On 32 RepoExec-inline tasks,
    \textsc{OpenCoder} improves GPT selected-output correctness by 21.88
    percentage points over plain RAG, while target-aware refinement
    substantially improves API-set retrieval. Additional stress testing
    characterizes how these benefits depend on the LLM backend and repository
    evidence completeness.
\end{itemize}

%Part2
\section{OpenCoder}
\label{sec:opencoder}
\subsection{Problem Formulation}
\label{subsec:problem-formulation}
Given a user query \(q\), consisting of a natural-language requirement and a
target function signature, repository-level code generation aims to produce an
implementation \(y\) that is consistent with the surrounding repository and
passes the associated validation tests. Let \(\mathcal{C}\) denote the
repository knowledge space. Following prior repository-level retrieval
settings~\cite{wu2024repoformer,gu2025what}, we organize \(\mathcal{C}\) into
three evidence sources:
\begin{equation}
\label{eq:knowledge-space}
\mathcal{C}=\mathcal{A}\cup\mathcal{X}\cup\mathcal{S},
\end{equation}
where \(\mathcal{A}\) contains project-specific API knowledge,
\(\mathcal{X}\) contains contextual repository code, and \(\mathcal{S}\)
contains semantically similar code examples.

For each source
\(m\in\{\mathcal{A},\mathcal{X},\mathcal{S}\}\), let
\(\mathcal{C}_m\subseteq\mathcal{C}\) be the corresponding source-specific
knowledge pool. A retriever \(\mathcal{R}_m\) returns source-specific evidence
\begin{equation}
\label{eq:source-retrieval}
E_m(q)=\mathcal{R}_m(q,\mathcal{C}_m),
\end{equation}
and the full retrieved evidence set is
\begin{equation}
\label{eq:evidence-union}
E(q)=E_{\mathcal{A}}(q)\cup E_{\mathcal{X}}(q)\cup E_{\mathcal{S}}(q).
\end{equation}
Conventional retrieval-augmented generation primarily ranks evidence according
to relevance. In contrast, \textsc{OpenCoder} additionally assigns each
retrieved item \(e\in E_m(q)\) a source-wise uncertainty score:
\begin{equation}
\label{eq:item-uncertainty}
u_m(e\mid q)\in[0,1],
\end{equation}
where larger values indicate lower confidence that the evidence is relevant,
complete, and compatible with the target implementation. The corresponding
source-level uncertainty is defined as
\begin{equation}
\label{eq:source-uncertainty}
\bar{u}_m(q)=
\frac{1}{|E_m(q)|}
\sum_{e\in E_m(q)}u_m(e\mid q).
\end{equation}

%To integrate heterogeneous evidence, \textsc{OpenCoder} computes an
%uncertainty-consensus score for each retrieved item:
%\begin{equation}
%\label{eq:uncertainty-consensus}
% IMPORTANT BEFORE PUBLIC RELEASE: replace \Gamma with the exact implemented
% aggregation rule and report its coefficients/normalization if available.
%\gamma(e\mid q,E)=
%\Gamma\!\left(
%r(e,q),\,
%1-u_m(e\mid q),\,
%a(e,E),\,
%c(e,E)
%\right),
%\end{equation}

%where \(r(e,q)\) denotes retrieval relevance, \(a(e,E)\) measures agreement
%with complementary evidence, and \(c(e,E)\) encodes redundancy or conflict;
%\(\Gamma\) decreases as the conflict term increases.

%Equation6 need to update again
To score retrieved evidence, \textsc{OpenCoder} combines relevance, predicted
source utility, and evidence uncertainty:
\begin{equation}
\label{eq:uncertainty-consensus}
\gamma_m(e\mid s)
=
r_m(e,q_{s,m})\,w_m(s)\,
\max\!\left\{0,1-\alpha u_m(e)\right\},
\end{equation}
where \(r_m\) is cosine similarity, \(w_m(s)\) is the normalized source-intent
weight, and \(u_m(e)\in[0,1]\) is evidence uncertainty. We set
\(\alpha=0.5\), retain the top \(70\%\) per source, and merge, deduplicate, and
rank the remaining evidence.
For each source, let \(\widetilde{E}_{s,m}\) contain the top \(70\%\) of
candidates ranked by \(\gamma_m(e\mid s)\). The final evidence set is
\begin{equation}
\label{eq:evidence-selection}
E^{*}(q)
=
\operatorname{TopK}_{\gamma}
\left(
\operatorname{Dedup}
\left(
\bigcup_{s,m}\widetilde{E}_{s,m}
\right)
\right),
\end{equation}
where \(K=10\) is the shared fused-context budget.

The generator produces candidate code conditioned on the query, selected
evidence, and source-wise uncertainty trace:
\begin{equation}
\label{eq:generation-distribution}
y\sim p_{\theta}
\left(
y\mid q,E^{*}(q),\mathbf{u}(q)
\right),
\end{equation}
where
\begin{equation}
\label{eq:uncertainty-trace}
\mathbf{u}(q)=
\left[
\bar{u}_{\mathcal{A}}(q),
\bar{u}_{\mathcal{X}}(q),
\bar{u}_{\mathcal{S}}(q)
\right]
\end{equation}
is the source-wise uncertainty trace. Given five candidates \(\mathcal{Y}(q)=(y_1,\ldots,y_5)\),
\textsc{OpenCoder} validates them in generation order and returns the first
passing candidate. If none passes, the earliest candidate
\(y_{\mathrm{mode}}\) from the largest normalized self-consistency group
undergoes at most two repair rounds:
\begin{equation}
\label{eq:candidate-selection}
\hat{y}
=
\begin{cases}
y_{j^{*}}, & j^{*}=\min\{j:\operatorname{Validate}(y_j)=1\},\\
\operatorname{Repair}^{(r^{*})}(y_{\mathrm{mode}}),
& r^{*}\leq 2 \text{ is the first passing repair},\\
\operatorname{Repair}^{(2)}(y_{\mathrm{mode}}), & \text{otherwise}.
\end{cases}
\end{equation}
Repair is triggered only when all raw candidates fail validation.

\subsection{Uncertainty-Aware Framework}
\label{subsec:framework}

As shown in Figure~\ref{fig:opencoder-framework}, OpenCoder is an
uncertainty-aware framework for retrieval-augmented repository-level code
generation. Unlike conventional RAG pipelines that mainly optimize retrieval
relevance, OpenCoder explicitly estimates, propagates, and mitigates
uncertainty across retrieval, generation, and repair. It comprises five phases. In Phase~I, it extracts repository APIs,
contextual code, and similar-code examples to construct an uncertainty-aware
knowledge index. Phase~II decomposes the query into implementation steps and
predicts the evidence required for each step. In Phase III, \textsc{OpenCoder} retrieves evidence from multiple sources,
scores candidates using retrieval relevance, predicted source utility, and
source-specific uncertainty, and then filters, deduplicates, and ranks the
retained evidence under a shared context budget. Phase~IV conditions the LLM on the selected evidence
and uncertainty trace, prioritizing reliable information while suppressing
noisy or incompatible context. Finally, Phase~V verifies generated candidates using static checks and
executable tests. The earliest passing candidate is selected; when no raw
candidate passes, the normalized-mode candidate undergoes at most two
validation-guided repair rounds.
\section{Experimental Setup}
\label{sec:experimental-setup}

\paragraph{Benchmarks and Task Selection.}
We select benchmarks according to two criteria: they should support executable
evaluation of functional correctness and require repository-specific context,
dependencies, or APIs. Accordingly, we use CoderEval, RepoExec, and
ExecRepoBench~\cite{yu2023codereval,hai2024repoexec,
yang2024execrepobench}. CoderEval contains 230 Python and 230 Java tasks collected from real-world
open-source projects and covers six levels of contextual dependency.
RepoExec evaluates repository-level function generation in terms of
executability, functional correctness, and dependency utilization.
ExecRepoBench contains approximately 1.2K Python repository-completion samples
constructed through AST-guided masking at statement, expression, and function
granularities. We adapt ten suitable
ExecRepoBench instances to function-level generation while preserving their
available repository context and executable tests. For RepoExec, we construct a controlled executable subset through a
pre-specified audit completed before examining any method outputs. Starting
from 14 validated \emph{RepoExec-inline} tasks, we assess the remaining 25
candidates and retain 18 that satisfy the frozen inclusion criteria:
dependency completeness, passing reference tests, and compatibility with the
evaluation harness. This yields 32 matched tasks for the final analysis. We
additionally use ten ExecRepoBench tasks as a deliberately context-limited
stress test to evaluate robustness under restricted repository evidence.

\paragraph{Evaluation Scope.}
The evaluation subset varies according to the artifacts required by each
research question. RQ1 and RQ2 use ten execution-backed ExecRepoBench tasks
with complete retrieval conditions, uncertainty traces, validation outcomes,
and repair records. RQ3 evaluates the expanded 32-task RepoExec-inline set and
the ten partial-context ExecRepoBench tasks. RQ4 uses 13 API-bearing RepoExec
tasks and 13 API-bearing CoderEval tasks for each LLM backend. The selected
ExecRepoBench tasks contain no resolvable repository-specific API calls and
are therefore excluded from API-set evaluation. All task-selection decisions
are completed before examining method outputs.

\paragraph{Compared Methods.}
Our primary controlled baseline, \emph{Baseline RAG}, retrieves API knowledge,
repository context, and similar-code evidence using the same retrieval and
candidate budgets as \textsc{OpenCoder}, but directly concatenates the
retrieved items with the generation prompt. It does not use uncertainty-aware
filtering, uncertainty-aware evidence integration, verification-guided selection, or repair.
\emph{RAG + Verify/Repair} receives the identical retrieved evidence as
Baseline RAG and applies the same executable validation and maximum two-round
repair budget as \textsc{OpenCoder}, but omits uncertainty-aware filtering, uncertainty-aware evidence
integration, and target-aware refinement. This matched control isolates
the contribution of uncertainty-aware evidence processing from downstream
verification and repair. For RQ4, \emph{OpenCoder w/o API Refinement} removes
target-aware API refinement while retaining the remaining pipeline.

\paragraph{LLM Backends and Generation Configuration.}
We evaluate \texttt{gpt-4o-mini} and \texttt{gemini-2.5-flash}. RQ1--RQ2 use
temperature \(0.2\) for controlled diagnostic analysis, whereas RQ3--RQ4 use
temperature \(0.7\) with five candidates per task and
\(k\in\{1,3,5\}\). RQ1 evaluates all \(2^3\) combinations of API knowledge,
repository context, and similar-code evidence across ten tasks and two
backends, yielding 160 task--condition runs and 480 executed generations.
RQ2 compares six paired component configurations. RQ3 evaluates Baseline RAG,
RAG + Verify/Repair, and \textsc{OpenCoder} on 32 RepoExec-inline tasks and
ten context-limited ExecRepoBench tasks using identical manifests, tests, and
a maximum of two repair rounds. The frozen protocol is applied without
retuning prompts, thresholds, or retrieval budgets.

\paragraph{Retrieval Configuration.}
We use UniXcoder~\cite{guo2022unixcoder} to embed code, natural-language
queries, and API descriptions in a shared dense space. Each source-specific
retriever returns up to eight candidates from API knowledge, repository
context, or similar-code evidence, and the fused context retains at most ten
items under a common prompt budget. \textsc{OpenCoder} ranks candidates using retrieval relevance, predicted
source utility, and source-specific uncertainty, filtering low-confidence
evidence before generation. Target-aware
refinement further removes self-retrieved, redundant, and
intent-incompatible APIs.

\paragraph{Uncertainty Estimation and Mitigation.}
\textsc{OpenCoder} estimates source-specific retrieval uncertainty for API,
repository-context, and similar-code evidence. Generation uncertainty combines
token entropy, self-consistency, and semantic variance with fixed weights
\(0.4\), \(0.4\), and \(0.2\) across all benchmarks and backends. These signals
guide evidence filtering and generation. Final candidate selection is
validation-driven: the earliest passing candidate is returned, and when no raw
candidate passes, the normalized-mode candidate undergoes at most two repair
rounds.

\paragraph{Evaluation Metrics.}
Functional correctness is measured using Pass@\(k\)
~\cite{chen2021evaluating}. For each task, we generate \(n\) candidate
implementations, of which \(c\) pass all executable tests, and compute
\begin{equation}
\mathrm{Pass@}k
=
1-\frac{\binom{n-c}{k}}{\binom{n}{k}}.
\label{eq:passk}
\end{equation}
We average the task-level estimates across benchmark instances and report
\(k\in\{1,3,5\}\) whenever \(n\geq k\). RepoExec-inline and the
partial-context ExecRepoBench setting use executable tests in RQ3; CoderEval
is excluded from this validated functional-correctness comparison. For RQ1, we report task-paired marginal effects of adding each retrieval source
on aggregate uncertainty and Pass@1, together with pairwise cross-source
interaction effects. For RQ2, we distinguish \emph{effective Pass@1}, which
evaluates the final verification-selected or repaired output, from
\emph{raw-sample Pass@1}, which measures the proportion of sampled candidates
that pass before selection and repair. We additionally report Expected
Calibration Error (ECE)~\cite{guo2017calibration}, failure-detection AUROC,
and the proportion of post-selection failures recovered through repair. For RQ3, we report candidate-set Pass@\(k\) and correctness of the single
selected output under a matched five-candidate budget. For RQ4, predicted and ground-truth API sets are compared using
macro-averaged precision, recall, F1, and exact API-set match. API-count
reliability is categorized as over-, exact-, or under-retrieval.   We further evaluate API-specific uncertainty using AUROC, AUPRC, and ECE on
a held-out task-level split, treating incorrectly retrieved API items as the
positive class.

\paragraph{Statistical Analysis.}
All evaluations use task-level paired comparisons. For RQ1, we estimate
marginal and pairwise interaction effects under the \(2^3\) factorial design,
with Holm correction within each backend and outcome family. For RQ2, changes
in binary correctness use two-sided exact McNemar tests with 95\%
paired-bootstrap confidence intervals. For RQ3, selected-output comparisons
use paired-bootstrap intervals and two-sided exact McNemar tests; candidate-set
Pass@\(k\) comparisons use paired-bootstrap intervals and exact paired tests.
The reported RQ3 McNemar values are nominal and unadjusted for multiple
comparisons, so we emphasize intervals and treat $p=.039$ as backend-specific
support rather than universal significance. Additional scope-specific results
are provided in the Technical Supplement. For RQ4, AUROC and ECE are computed
exclusively on a held-out task split.
\section{Results}
\label{sec:results}

\subsection{RQ1: Influence of Retrieved Evidence}
\label{subsec:rq1-results}
\begin{figure*}[t]
\centering
\begin{subfigure}[t]{0.49\textwidth}
    \centering
    \includegraphics[width=\linewidth]{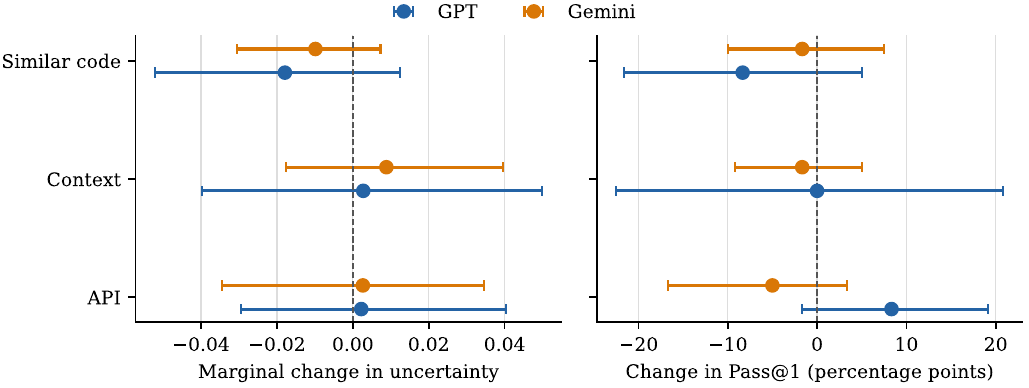}
    \caption{Task-paired marginal effects of adding each evidence source.}
    \label{fig:source-marginal}
\end{subfigure}
\hfill
\begin{subfigure}[t]{0.49\textwidth}
    \centering
    \includegraphics[width=\linewidth]{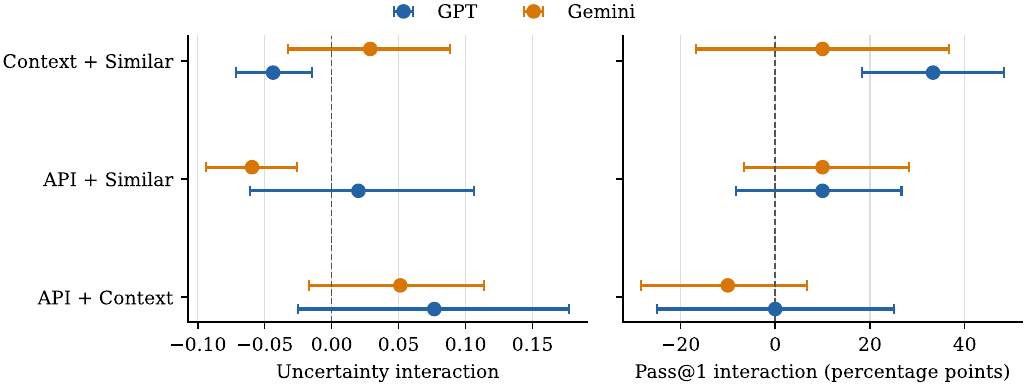}
    \caption{Task-paired pairwise interactions between evidence sources.}
    \label{fig:source-interactions}
\end{subfigure}
%\vspace{-5pt}
\caption{Factorial effects of API knowledge, repository context, and
similar-code evidence on aggregate uncertainty and Pass@1. Points show
task-paired effect estimates, and error bars denote 95\% bootstrap confidence
intervals. Positive values indicate increased uncertainty ($\Delta U$) or
improved functional correctness (Pass@1); significance is Holm-adjusted.}
\label{fig:retrieval-evidence-effects}
\end{figure*}

We evaluate all \(2^3\) combinations of API knowledge, repository context, and
similar-code evidence on ten execution-backed ExecRepoBench tasks for each LLM
backend. This factorial evaluation comprises 160 task--condition runs and 480
test-executed candidate generations. Table~\ref{tab:rq1-factorial} reports the
exact task-paired marginal effects, while
Figure~\ref{fig:retrieval-evidence-effects} visualizes both the marginal and
pairwise interaction effects with 95\% bootstrap confidence intervals. After Holm
correction, no individual source exhibits a statistically significant marginal
effect on either aggregate uncertainty or Pass@1.

The strongest effects instead arise from interactions among evidence sources.
For GPT, repository context and similar-code evidence exhibit a positive
Pass@1 interaction of 33.3 percentage points
(\(p_{\mathrm{Holm}}=.032\)). For Gemini, combining API and similar-code
evidence reduces aggregate uncertainty by \(0.059\)
(\(p_{\mathrm{Holm}}=.032\)). These results indicate that the contribution of
a retrieval source depends on both the accompanying evidence and the LLM
backend, rather than following a universal additive ranking.

\begin{table}[t]
\centering

\scriptsize
\resizebox{\columnwidth}{!}{%
\begin{tabular}{llrr}
\toprule
LLM & Evidence source
    & \(\Delta U\) (\(p_{\mathrm{H}}\))
    & \(\Delta\)Pass@1 (\(p_{\mathrm{H}}\)) \\
\midrule
GPT & API knowledge      & \(+0.002\) (1.000) & \(+8.3\) (.623) \\
GPT & Repository context & \(+0.003\) (1.000) & \(0.0\) (.984) \\
GPT & Similar code       & \(-0.018\) (1.000) & \(-8.3\) (.623) \\
\midrule
Gemini & API knowledge      & \(+0.003\) (1.000) & \(-5.0\) (1.000) \\
Gemini & Repository context & \(+0.009\) (1.000) & \(-1.7\) (1.000) \\
Gemini & Similar code       & \(-0.010\) (1.000) & \(-1.7\) (1.000) \\
\bottomrule
\end{tabular}%
}
\vspace{-5pt}
\caption{Task-paired marginal effects of adding each retrieval source.
Parentheses contain Holm-adjusted \(p\)-values. Pass@1 effects are reported in
percentage points.}
\label{tab:rq1-factorial}
\end{table}

\noindent
\fbox{%
\parbox{\dimexpr\columnwidth-2\fboxsep-2\fboxrule\relax}{%
\small
\textbf{Finding 1.}
Retrieval utility is interaction-dependent: the contribution of API knowledge,
repository context, and similar-code evidence varies with their combination
and the LLM backend. This finding empirically motivates uncertainty-aware multi-source integration over fixed, source-wise weighting.
}}

\subsection{RQ2: Uncertainty Quantification and Mitigation}
\label{subsec:rq2-results}

\begin{figure}[t]
\centering
\includegraphics[width=\columnwidth]{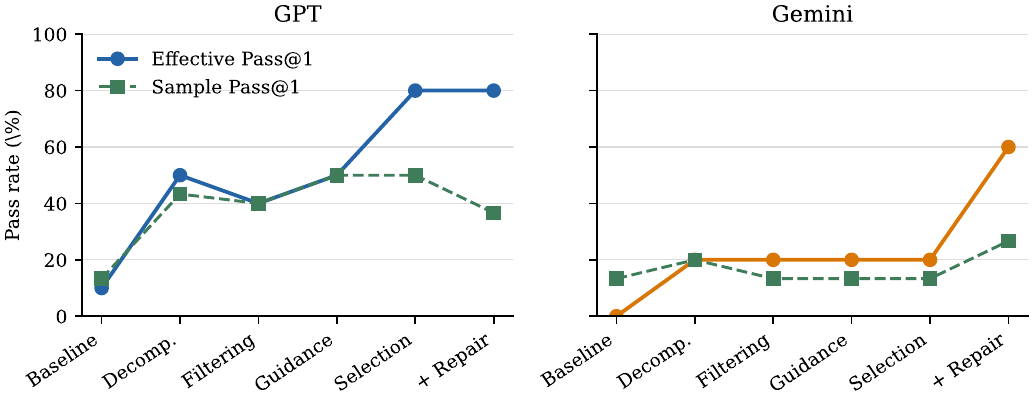}
%\vspace{-15pt}
\caption{Effective and raw-sample Pass@1 as uncertainty-aware components are
introduced. Effective Pass@1 evaluates the final selected or repaired output,
whereas raw-sample Pass@1 measures correctness before verification-based
selection and repair.}
\label{fig:component-effectiveness}
\end{figure}

We isolate the effects of query decomposition, uncertainty filtering, guided
generation, verified selection, and repair using six paired configurations for
each task and LLM backend. Figure~\ref{fig:component-effectiveness}
distinguishes \emph{effective Pass@1}, which evaluates the final selected or
repaired output, from correctness measured over the three raw candidate
samples. On the ten-task diagnostic subset, the complete \textsc{OpenCoder}
pipeline increases
effective Pass@1 from 10.0\% to 80.0\% for GPT
(95\% CI [40.0, 100.0], exact McNemar \(p=.016\)) and from 0.0\% to
60.0\% for Gemini (95\% CI [30.0, 90.0], \(p=.031\)). The component analysis
attributes most of the GPT improvement to verification-based candidate
selection. For Gemini, the repair stage recovers 33.3\% of generations that
remain incorrect after selection.
Uncertainty is useful as a control signal, but its reliability is strongly
backend-dependent. ECE changes from 0.162 to 0.222 for GPT and from 0.191 to
0.153 for Gemini. Failure-detection AUROC is 0.167 for GPT, indicating an
inverse association under the evaluated score orientation, but 0.905 for
Gemini. We therefore use uncertainty operationally to control evidence
selection, validation, and repair rather than assuming backend-independent
probabilistic calibration.

\noindent
\fbox{%
\parbox{\dimexpr\columnwidth-2\fboxsep-2\fboxrule\relax}{%
\small
\textbf{Finding 2: Uncertainty becomes effective through actionable control.}
By combining uncertainty-aware evidence filtering with executable verification
and repair, \textsc{OpenCoder} substantially improves selected-output
correctness on the diagnostic subset. Differences across LLM backends
underscore the importance of backend-adaptive uncertainty interpretation.
}}

\subsection{RQ3: End-to-End Effectiveness}
\label{subsec:rq3-results}
\vspace{-5pt}
Prior to generation, we audit the remaining 25 RepoExec tasks under a frozen
executable protocol and retain 18, yielding an expanded set of 32 matched
RepoExec-inline tasks. We evaluate all three methods on this set and on a
ten-task partial-context ExecRepoBench stress test. Table~\ref{tab:rq3-end-to-end} reports the expanded results together with
the ten-task partial-context ExecRepoBench stress test. All methods use
identical task manifests, retrieval evidence, temperature, five-candidate
budgets, executable tests, and at most two repair rounds.

\begin{table*}[t]
\centering

{\footnotesize
\setlength{\tabcolsep}{3pt}
\renewcommand{\arraystretch}{1.0}

\begin{adjustbox}{max width=\textwidth}
\begin{tabular}{@{}llcccccccc@{}}
\toprule
\multirow{2}{*}{LLM} & \multirow{2}{*}{Method}
& \multicolumn{4}{c}{RepoExec-inline ($N=32$)}
& \multicolumn{4}{c}{ExecRepoBench$^{\dagger}$ ($N=10$)} \\
\cmidrule(lr){3-6}\cmidrule(lr){7-10}
& & Pass@1 & Pass@3 & Pass@5 & Sel.
  & Pass@1 & Pass@3 & Pass@5 & Sel. \\
\midrule
GPT & Baseline RAG
& 58.75 & 68.44 & 71.88 & 56.25
& 52.00 & 82.00 & \textbf{100.00} & 40.00 \\

GPT & RAG + Verify/Repair
& \textbf{61.88} & \textbf{74.38} & \textbf{78.13} & \textbf{78.13}
& \textbf{62.00} & \textbf{95.00} & \textbf{100.00} & \textbf{100.00} \\

GPT & \textsc{OpenCoder}
& \textbf{61.88} & 72.50 & \textbf{78.13} & \textbf{78.13}
& 28.00 & 39.00 & 40.00 & 40.00 \\
\midrule
Gemini & Baseline RAG
& 66.25 & 70.31 & 71.88 & 68.75
& 90.00 & 99.00 & \textbf{100.00} & 80.00 \\

Gemini & RAG + Verify/Repair
& \textbf{70.00} & \textbf{78.75} & \textbf{84.38} & \textbf{84.38}
& \textbf{94.00} & \textbf{100.00} & \textbf{100.00} & \textbf{100.00} \\

Gemini & \textsc{OpenCoder}
& 65.00 & 73.75 & 78.13 & 78.13
& 48.00 & 68.00 & 80.00 & 80.00 \\
\bottomrule
\end{tabular}
\end{adjustbox}
}
%\vspace{-5pt}
\caption{End-to-end RQ3 results under the frozen five-candidate protocol.
ExecRepoBench$^{\dagger}$ is evaluated under partial repository context.
Sel. denotes selected-output correctness; bold marks the best result per
LLM and benchmark, including ties.}
\label{tab:rq3-end-to-end}
\end{table*}

With GPT, \textsc{OpenCoder} improves RepoExec-inline selected-output
correctness over Baseline RAG from 56.25\% to 78.13\% (\(\Delta=+21.88\),
95\% CI [6.25,40.63], W/L/T $=8/1/23$, nominal McNemar $p=.039$), but ties
RAG+Verify/Repair at 78.13\% (\(\Delta=0\), CI [-9.38,9.38]). Its
Pass@1/3/5 values are 61.88/72.50/78.13, compared with
61.88/74.38/78.13 for the matched control. Thus, the expanded experiment
provides backend-specific evidence for better GPT output selection than plain
RAG, but not for an advantage beyond executable verification and repair. With Gemini, \textsc{OpenCoder} obtains selected-output correctness of 78.13\%,
compared with 68.75\% for Baseline RAG and 84.38\% for RAG+Verify/Repair. The
paired differences are not statistically supported (Baseline: \(\Delta=+9.38\),
CI [-6.25,25.00], $p=.453$; control: \(\Delta=-6.25\), CI [-15.63,0],
$p=.500$). The control is also strongest across Pass@1/3/5. Across both
backends, all paired RepoExec-inline Pass@\(k\) confidence intervals include
zero. The partial-context ExecRepoBench results retain the previously observed
boundary condition: RAG+Verify/Repair outperforms \textsc{OpenCoder} at every
candidate-set metric and in selected-output correctness. Incomplete repository
evidence can therefore make uncertainty-aware filtering suppress useful
alternatives rather than improve the final decision.

\noindent
\fbox{%
\parbox{\dimexpr\columnwidth-2\fboxsep-2\fboxrule\relax}{%
\small
\textbf{Finding 3: Uncertainty-aware control strengthens selected-output
reliability when repository evidence is sufficient.}
On the expanded 32-task RepoExec-inline set, \textsc{OpenCoder} improves GPT
selected-output correctness by 21.88 percentage points over Baseline RAG and
matches the verification-and-repair control. Results across backends and the
context-limited stress test further show that these benefits depend on both
LLM behavior and repository-evidence completeness.
}}

\subsection{RQ4: API Retrieval Reliability}
\label{subsec:rq4-results}
\begin{figure}[h]
\centering
\includegraphics[width=\columnwidth]{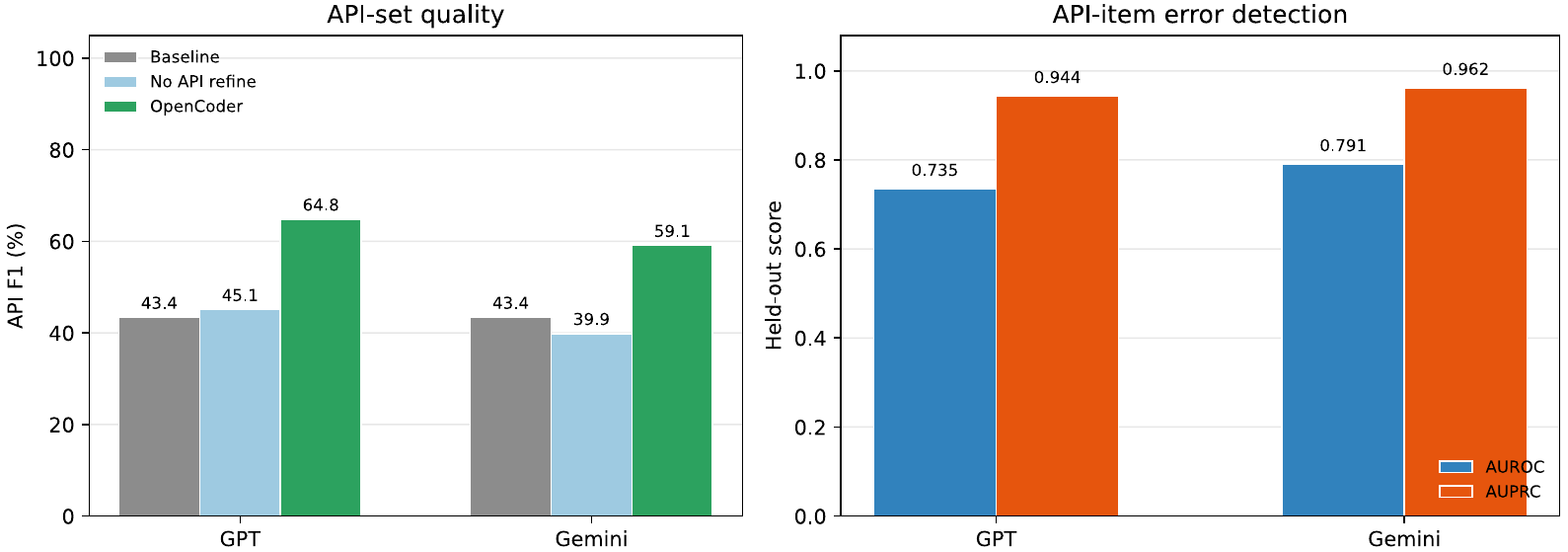}
%\vspace{-15pt}
\caption{API-set retrieval quality (left) and held-out false-positive API
detection (right).}
\label{fig:api-reliability}
\end{figure}

We evaluate API retrieval on 13 API-bearing RepoExec and 13 API-bearing
CoderEval tasks for each LLM backend. The selected ExecRepoBench tasks contain
no resolvable repository-specific API calls and are therefore excluded from
the API-F1 analysis. As shown in Figure~\ref{fig:api-reliability}, target-aware refinement
increases macro-averaged API F1 from 43.4\% to 64.8\% for GPT and from
43.4\% to 59.1\% for Gemini. Exact API-set match increases from 0.0\% to
57.7\% and 50.0\%, respectively. Removing API refinement reduces F1 to
45.1\% for GPT and 39.9\% for Gemini, identifying refinement as the primary
contributor to the improvement. API-count reliability nevertheless varies substantially across benchmarks.
\textsc{OpenCoder} retrieves the exact number of required APIs for 100.0\% of
GPT and 92.9\% of Gemini RepoExec tasks, but for only 10.5\% of CoderEval
tasks. It over-retrieves APIs for the remaining 89.5\% of CoderEval tasks. On a held-out task split, API-specific uncertainty detects
incorrectly retrieved API items with AUROC values of 0.735 and 0.791 and ECE
values of 0.030 and 0.041 for GPT and Gemini, respectively. This signal can
therefore support filtering of false-positive evidence, but cannot identify
required APIs that are absent from the retrieved candidate set. Exact API
recovery is also insufficient for functional correctness: although all
13 GPT RepoExec tasks recover the exact API set, only ten generated
implementations pass their executable tests.

\begin{table}[t]
\centering
{\small
\setlength{\tabcolsep}{4.5pt}
\renewcommand{\arraystretch}{1.0}

\textbf{(a) API-set retrieval quality (\%)}\\[-2pt]
\begin{tabular}{@{}llrr@{}}
\toprule
LLM & Method & Macro F1 & Exact \\
\midrule
\multirow{3}{*}{GPT}
& Baseline RAG
& 43.4 & 0.0 \\
& \textsc{OpenCoder} w/o API refinement
& 45.1 & 3.8 \\
& \textsc{OpenCoder}
& \textbf{64.8} & \textbf{57.7} \\
\addlinespace[1pt]
\multirow{3}{*}{Gemini}
& Baseline RAG
& 43.4 & 0.0 \\
& \textsc{OpenCoder} w/o API refinement
& 39.9 & 0.0 \\
& \textsc{OpenCoder}
& \textbf{59.1} & \textbf{50.0} \\
\bottomrule
\end{tabular}

\vspace{6pt}

\textbf{(b) False-positive API detection}\\[-2pt]
\begin{tabular}{@{}lrrr@{}}
\toprule
LLM & AUROC & AUPRC & ECE \\
\midrule
GPT    & 0.735 & 0.944 & 0.030 \\
Gemini & 0.791 & 0.962 & 0.041 \\
\bottomrule
\end{tabular}
}
%\vspace{-5pt}
\caption{API retrieval quality and false-positive detection.
Exact denotes exact API-set recovery over 26 API-bearing tasks per backend:
13 RepoExec and 13 CoderEval tasks.}
\label{tab:api-reliability}
\end{table}

Table~\ref{tab:api-reliability} shows that target-aware refinement is critical
for API grounding. It raises macro F1 from 45.1 to 64.8 for GPT and from 39.9
to 59.1 for Gemini, while exact-set recovery increases from 3.8\% to 57.7\%
and from 0.0\% to 50.0\%, respectively. API-specific uncertainty further
detects false-positive evidence with AUROC values of 0.735 and 0.791 and low
ECE, supporting uncertainty-guided filtering.

\noindent
\fbox{%
\parbox{\dimexpr\columnwidth-2\fboxsep-2\fboxrule\relax}{%
\small
\textbf{Finding 4: Target-aware refinement strengthens API grounding.}
\textsc{OpenCoder} improves API-set quality through target-aware refinement,
while API-specific uncertainty provides an actionable signal for identifying
false-positive evidence. Together, these mechanisms enhance repository
grounding, with recovery of omitted APIs depending on adequate candidate
coverage.
}}

\section{Related Work}
\label{sec:related-work}

\paragraph{Retrieval-Augmented Code Generation.}
Retrieval-augmented code generation uses related code, documentation, and API
knowledge to supplement a model's internal knowledge
~\cite{hayati2018recode,parvez2021redcoder,lu2022reacc,
zhou2022docprompting,zan2023apicoder}. Repository-level methods extend this
paradigm to cross-file dependencies, project conventions, and private APIs.
RepoCoder alternates retrieval and generation, whereas RepoFormer and SRACG
selectively invoke or filter retrieval to reduce harmful augmentation
~\cite{zhang2023repocoder,wu2024repoformer,wang2026sracg}.
Other systems improve context construction through multi-source fusion,
structural retrieval, and learned selection
~\cite{liang2024repofuse,liao2024a3codgen,liu2024graphcoder,
deng2024r2c2coder,wang2024rlcoder}. Most closely related, Gu et al.\ compare
API knowledge, repository context, and similar-code evidence and demonstrate
that their utility varies across source types~\cite{gu2025what}.
Unlike approaches centered primarily on retrieval effectiveness,
\textsc{OpenCoder} explicitly models evidence reliability and uses it
throughout generation and validation.
\paragraph{Uncertainty and Execution-Guided Reliability.}
LLM uncertainty has been studied through semantic consistency, uncertainty
decomposition, and confidence calibration
~\cite{farquhar2024semanticentropy,hou2024decomposing,
shen2024thermometer}. In code generation, Incoherence estimates error from
behavioral disagreement without an execution oracle
~\cite{valentin2026incoherence}. CodeT and LEVER use generated tests or
execution outcomes for candidate selection, while self-debugging iteratively
repairs incorrect programs
~\cite{chen2023codet,ni2023lever,chen2024selfdebug}.
These approaches primarily assess or correct uncertainty after generation;
\textsc{OpenCoder} additionally models uncertainty in the repository evidence
used to construct the generation context.
\paragraph{Adaptive and Reliability-Aware Retrieval.}
Self-RAG and CRAG determine when evidence should be retrieved, critiqued, or
corrected~\cite{asai2023selfrag,yan2024crag}. Probing-RAG predicts retrieval
necessity, while Reliability-Aware RAG estimates heterogeneous source
reliability to prioritize trustworthy evidence
~\cite{baek2024probingrag,hwang2024rarag}. These methods establish retrieval
necessity and source reliability as useful control signals, but primarily
target knowledge-intensive text generation. \textsc{OpenCoder} specializes
these ideas for repository-level code generation by estimating uncertainty
over similar code, repository context, and API evidence, integrating the sources through uncertainty-aware ranking, and coupling them with executable
verification and repair.

%\vspace{-5pt}
\section{Limitations and Ethical Considerations}
\label{sec:limitations}

Our validated evaluation covers 32 RepoExec-inline tasks and a ten-task
partial-context ExecRepoBench stress test. The GPT selected-output gain over
Baseline RAG is the only comparison with an unadjusted confidence interval
excluding zero (nominal $p=.039$); the method matches the
verification-and-repair control, while the corresponding Gemini and
candidate-set Pass@\(k\) differences are not statistically supported.
Generalization may therefore depend on the backend, benchmark scope, and
repository-evidence completeness. The framework also incurs additional
inference cost, and its uncertainty estimates are not universally calibrated.
Required APIs absent from the initial candidate pool cannot be recovered through
filtering alone. Generated code may still contain functional, security, or
licensing defects and should undergo developer review and project-specific
testing before deployment.
\section{Conclusion}
\label{sec:conclusion}

We introduced \textsc{OpenCoder}, an uncertainty-aware framework that jointly
controls heterogeneous repository evidence, generation, verification, and
repair. Our factorial analysis shows that retrieval utility emerges from
cross-source interactions, motivating uncertainty-aware multi-source integration rather than fixed
source weighting. On 32 pre-audited RepoExec-inline tasks, \textsc{OpenCoder}
improves GPT selected-output correctness by 21.88 percentage points over
Baseline RAG and matches RAG+Verify/Repair under the same validation-and-repair
budget. Target-aware API refinement further improves API-set grounding, while
results with Gemini and under partial repository context show that the benefits
depend on backend behavior and evidence completeness. Overall, uncertainty is
most useful not as an isolated confidence estimate, but as an operational
control signal coupled with evidence fusion, executable validation, and repair.
\bibliography{aaai2027}
\end{document}